\documentclass[aps,preprint,showpacs,groupedaddress]{revtex4-1}
\usepackage{amssymb}
\usepackage{mathrsfs}
\usepackage{amsmath}
\usepackage{mathtools}
\usepackage{pstricks}
\usepackage{color}
\usepackage{graphicx}
\usepackage{amsthm}

\usepackage{physics}

\begin{document}


\title{\bf Influence of exact Lorentz-violating mechanism on the critical exponents for massive $q$-deformed $\lambda\phi^{4}$ scalar field theory}



\author{P. R. S. Carvalho}
\email{prscarvalho@ufpi.edu.br}
\affiliation{\it Departamento de F\'\i sica, Universidade Federal do Piau\'\i, 64049-550, Teresina, PI, Brazil}

\author{M. I. Sena-Junior}
\email{marconesena@poli.br}
\affiliation{\it Escola Polit\'{e}cnica de Pernambuco, Universidade de Pernambuco, 50720-001, Recife, PE, Brazil}




\begin{abstract}
We probe the influence of Lorentz-violating mechanism, treated exactly, on the radiative quantum corrections to critical exponents for massive $q$-deformed O($N$) $\lambda\phi^{4}$ scalar field theories. We attain that task by employing three distinct and independent field-theoretic renormalization group methods. Firstly we compute the critical exponents up to the finite next-to-leading order for then generalizing the results for any loop level. We show that the $q$-deformed critical exponents are insensible to the Lorentz symmetry breaking mechanism thus agreeing with the universality hypothesis.  
\end{abstract}


\maketitle


\section{Introduction}\label{Introduction} 

\par A cosmological constant term in $3d$ quantum gravity can be produced by a $q$-deformation symmetry mechanism \cite{Livine2017}, as well as in others gravity theories \cite{Dil,1751-8121-42-42-425402,0264-9381-24-13-009,MAJOR1996267}. The knowledge of the symmetry properties of a given physical system is a very important feature to consider in describing its behavior. For example, in the high energy physics realm, three of the four elementary interactions, \emph{i. e.} the electromagnetic, weak and strong interactions, are described by the standard model of elementary particles and fields which is invariant under gauge symmetries \cite{PeskinSchroeder,ItzyksonZuber}. The same situation is observed when we are dealing with low energy physics problems, specifically in studying the critical behavior of systems undergoing a continuous phase transition \cite{Stanley}. The latter systems present a scaling critical behavior which is characterized by a set of critical exponents. These critical indices are universal quantities and depend on a few universal properties of the systems as their dimension $d$, $N$ and symmetry of some order parameter (the magnetization for magnetic systems) if the interactions of its constituents are of short- or long-range type and not on nonuniversal properties as their critical temperature or form of their lattices. Surprisingly, many distinct systems as a fluid and a ferromagnet can display an identical set of critical exponents if they share the same universal properties just mentioned. When this happens, we say that these distinct systems belong to the same universality class. This is, in essence, the content of the universality hypothesis. The universality class approached in this work is the O($N$) one, which is reduced to the cases where short-range interactions are present, namely the Ising ($N=1$), XY ($N=2$), Heisenberg ($N=3$), self-avoiding random walk ($N=0$), spherical ($N \rightarrow \infty$) models etc \cite{Pelissetto2002549}. 

\par Recently, the radiative quantum corrections to critical exponents for $q$-deformed O($N$) $\lambda\phi^{4}$ scalar field theories were evaluated , at least, at next-to-leading order \cite{PhysRevD.97.105006}. This theory is a CPT-even aether-like scalar field theory \cite{PhysRevD.58.116002,PhysRevD.78.044047,PhysRevD.81.045018}, for which previous results were obtained as its insertion in the Lorentz-violating extension of the standard model, dispersion relations, aether compactification, the one-loop effective potential corresponding to the action of the scalar field etc. In fact, $q$-deformed systems have attracted great attention in the last few years \cite{Adv.High.EnergyPhys.20179530874,Eur.Phys.J.Plus1322017398,Int.J.Theor.Phys.5620171746,Phys.DarkUniv.1620171,Adv.HighEnergyPhys.20179371391,EPL11320162000,JHEP1220160630,Phys.Rev.D912015044024,G.Vinod}. The aim of the present work is to probe the influence of the Lorentz symmetry breaking effect \cite{Universe2201630,PhysRevD.39.683,Kostelecký1991545,PhysRevD.59.124021,Amelino-Camelia20135,PhysRevLett.90.211601,Hayakawa200039,PhysRevLett.87.141601,PhysRevD.92.045016,doi:10.1142/S0217751X15500724,VANTILBURG2015236,PhysRevD.86.125015,PhysRevD.84.065030,Carvalho2014320,Carvalho2013850}, treated exactly \cite{PhysRevLett.83.2518,PhysRevD.96.116002,CARVALHO2017290,Carvalho2017}, on the $q$-deformed critical exponents values. For that, we employ thee distinct and independent field-theoretic renormalization group methods for renormalizing a massive theory. As the critical exponents are universal quantities, they must be the same when obtained through any method. In this field-theoretic approach, a massive theory corresponds to a system near but not at the critical temperature $T = T_{c}$. In this case, we have to renormalize a massive field theory. This renormalization procedure is needed since the theory in its initially bare form is plagued by ultraviolet divergences. The divergences are a result of the commutation relations among the fields when computed at the same point of spacetime    
\begin{eqnarray}\label{kljkkjij}
[a(k)_{q}, a_{q}^{\dagger}(k^{\prime})] \equiv a(k)_{q}a_{q}^{\dagger}(k^{\prime}) - q^{-1}a_{q}^{\dagger}(k^{\prime})a(k)_{q} = q^{N(k)}\delta(k - k^{\prime}), 
\end{eqnarray}  
\begin{eqnarray}\label{sfdfgcvcgf}
[a(k)_{q}, a_{q}(k^{\prime})] = 0, 
\end{eqnarray} 
\begin{eqnarray}\label{sfcvcgf}
[a^{\dagger}(k)_{q}, a_{q}^{\dagger}(k^{\prime})] = 0, 
\end{eqnarray}    
where $N(k) = a_{q}^{\dagger}(k)a_{q}(k)$ is the $q$-deformed number operator and
\begin{eqnarray}\label{ygfdxzsze}
&&\phi_{q}(x) = \int \frac{d^{3}k}{(2\pi)^{3/2}2\omega_{k}^{1/2}}[a(k)_{q}\exp^{-ikx} +a_{q}^{\dagger}(k)\exp^{ikx}]\nonumber \\ &&
\end{eqnarray}
where $\omega_{k}^{2} = \vec{k}^{2} + m^{2}$ whose bare free propagator is given by $G_{0B}(k) = \parbox{12mm}{\includegraphics[scale=1.0]{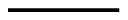}} = q/(k^{2} + K_{\mu\nu}k^{\mu}k^{\nu} + m_{B}^{2})$ \cite{G.Vinod}. The effect of these divergences is shown in the divergent scaling behavior of the $1$PI vertex parts \cite{Amit} thus giving rise to the emergence of anomalous dimensions. These anomalous dimensions furnish two of the six critical exponents we have to compute when evaluated at the nontrivial fixed point. As there are four scaling relations among the critical exponents, we must evaluate only two of them independently \cite{Stanley}. We have to calculate the $\eta_{q}$ and $\nu_{q}$ Lorentz-violating $q$-deformed critical exponents and the remaining ones through the scaling relations. The Lorentz-violating $q$-deformed nontrivial fixed point is obtained as the nontrivial root of the Lorentz-violating $q$-deformed $\beta_{q}$-function. It is responsible to take into account the radiative quantum corrected Lorentz-violating $q$-deformed critical exponents, while the trivial fixed point is obtained as the trivial solution of $\beta_{q}$ and would turn us able to compute the indices only in the mean field or Landau approximation \cite{Stanley}. 

\par This work proceeds as follows: Firstly, we have to renormalize the theory in three different and independent renormalization schemes and compute the corresponding critical exponents, at least, at next-to-leading order. Secondly we have to generalize the results from next-to-leading order for any loop level. At the end we present our conclusions.

\section{Renormalization group methods for massive theories}\label{Renormalization group methods for massive theories}

\par Now we have to approach the problem of computing the Lorentz-violating $q$-deformed critical exponents through the three methods displayed below.

\subsection{Bogoliubov-Parasyuk-Hepp-Zimmermann method}

\par The first method to be applied is the BPHZ (Bogoliubov-Parasyuk-Hepp-Zimmermann) one \cite{BogoliubovParasyuk,Hepp,Zimmermann}. It is the most general method used here. It is based on the Feynman diagrams evaluated at arbitrary external momenta values and include a maximum number of diagrams and counterterms as opposed to the two another distinct methods to be displayed later. The divergences of the theory are absorbed by the renormalization constants \cite{Kleinert} 
\begin{eqnarray}\label{Zphi}
&& Z_{\phi}(u,\epsilon^{-1}) = 1 +  \frac{1}{P^{2} + K_{\mu\nu}P^{\mu}P^{\nu}} \Biggl[ \frac{1}{6} \mathcal{K} 
\left(\parbox{12mm}{\includegraphics[scale=1.0]{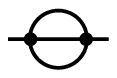}}
\right) \Biggr|_{m^2 = 0} S_{\parbox{10mm}{\includegraphics[scale=0.5]{fig6.eps}}} +  \frac{1}{4} \mathcal{K} 
\left(\parbox{12mm}{\includegraphics[scale=1.0]{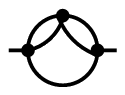}} \right) \Biggr|_{m^2 = 0} S_{\parbox{6mm}{\includegraphics[scale=0.5]{fig7.eps}}} + \nonumber \\&& \frac{1}{3} \mathcal{K}
  \left(\parbox{12mm}{\includegraphics[scale=1.0]{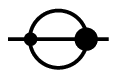}} \right) S_{\parbox{6mm}{\includegraphics[scale=0.5]{fig26.eps}}} \Biggr], \hspace{4mm}
\end{eqnarray}

\begin{eqnarray}\label{Zg}
&&Z_{u}(u,\epsilon^{-1}) = 1 + \frac{1}{\mu^{\epsilon}u} \Biggl[ \frac{1}{2} \mathcal{K} 
\left(\parbox{10mm}{\includegraphics[scale=1.0]{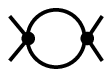}} + 2 \hspace{1mm} perm.
\right) S_{\parbox{10mm}{\includegraphics[scale=0.5]{fig10.eps}}} + \frac{1}{4} \mathcal{K} 
\left(\parbox{17mm}{\includegraphics[scale=1.0]{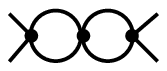}} + 2 \hspace{1mm} perm. \right) S_{\parbox{10mm}{\includegraphics[scale=0.5]{fig11.eps}}} + \nonumber \\ && \frac{1}{2} \mathcal{K} 
\left(\parbox{12mm}{\includegraphics[scale=.8]{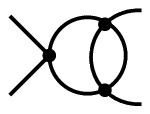}} + 5 \hspace{1mm} perm. \right) S_{\parbox{10mm}{\includegraphics[scale=0.4]{fig21.eps}}} + \frac{1}{2} \mathcal{K} 
\left(\parbox{10mm}{\includegraphics[scale=1.0]{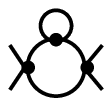}} + 2 \hspace{1mm} perm.
\right) S_{\parbox{10mm}{\includegraphics[scale=0.5]{fig13.eps}}} + \nonumber \\ && \mathcal{K}
  \left(\parbox{10mm}{\includegraphics[scale=1.0]{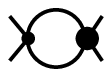}} + 2 \hspace{1mm} perm. \right) S_{\parbox{6mm}{\includegraphics[scale=0.5]{fig25.eps}}} + \mathcal{K}
  \left(\parbox{10mm}{\includegraphics[scale=1.0]{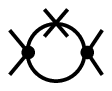}} + 2 \hspace{1mm} perm. \right) S_{\parbox{6mm}{\includegraphics[scale=0.5]{fig24.eps}}} \Biggr],
\end{eqnarray}

\begin{eqnarray}\label{Zphi}
&& Z_{m^{2}}(u,\epsilon^{-1}) = 1 + \frac{1}{m^{2}} \Biggl[ \frac{1}{2} \mathcal{K} 
\left(\parbox{12mm}{\includegraphics[scale=1.0]{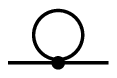}}
\right) S_{\parbox{10mm}{\includegraphics[scale=0.5]{fig1.eps}}} + \frac{1}{4} \mathcal{K} 
\left(\parbox{12mm}{\includegraphics[scale=1.0]{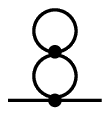}} \right) S_{\parbox{6mm}{\includegraphics[scale=0.5]{fig2.eps}}} + \frac{1}{2} \mathcal{K}
  \left(\parbox{12mm}{\includegraphics[scale=1.0]{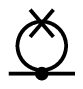}} \right) S_{\parbox{6mm}{\includegraphics[scale=0.5]{fig22.eps}}} + \nonumber \\ && \frac{1}{2} \mathcal{K}
  \left(\parbox{12mm}{\includegraphics[scale=1.0]{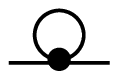}} \right) S_{\parbox{6mm}{\includegraphics[scale=0.5]{fig23.eps}}} + \frac{1}{6} \mathcal{K}
  \left(\parbox{12mm}{\includegraphics[scale=1.0]{fig6.eps}} \right)\Biggr|_{P^2 + K_{\mu\nu}P^{\mu}P^{\nu} = 0} S_{\parbox{6mm}{\includegraphics[scale=0.5]{fig6.eps}}} \Biggr], \hspace{4mm}
\end{eqnarray}
where we have that $S_{\parbox{6mm}{\includegraphics[scale=0.5]{fig6.eps}}}$, for example, is the symmetry factor for that diagram when for a field with $N$ components. The evaluated diagrams and counterterms are given by
\begin{eqnarray}
\left(\parbox{12mm}{\includegraphics[scale=1.0]{fig6.eps}}\right)\Biggr|_{m^{2}=0} =  -\frac{u^{2}(P^2 + K_{\mu\nu}P^{\mu}P^{\nu})}{8\epsilon} \left[ 1 + \frac{1}{4}\epsilon -2\epsilon\, J_{3}(P^{2} + K_{\mu\nu}P^{\mu}P^{\nu}) \right]q^{3}\mathbf{\Pi}^{2},
\end{eqnarray}
\begin{eqnarray}
\parbox{12mm}{\includegraphics[scale=1.0]{fig7.eps}}\bigg|_{m^{2}=0} = \frac{(P^2 + K_{\mu\nu}P^{\mu}P^{\nu})u^{3}}{6\epsilon^{2}} \left[1 + \frac{1}{2}\epsilon - 3\epsilon\, J_{3}(P^{2} + K_{\mu\nu}P^{\mu}P^{\nu})\right]q^{5}\mathbf{\Pi}^{3},
\end{eqnarray}
\begin{eqnarray}
\parbox{10mm}{\includegraphics[scale=1.0]{fig26.eps}} \quad = -\frac{3(P^{2} + K_{\mu\nu}P^{\mu}P^{\nu})u^{3}}{16\epsilon^{2}}\left[1 + \frac{1}{4}\epsilon - 2\epsilon\, J_{3}(P^{2} + K_{\mu\nu}P^{\mu}P^{\nu})\right]q^{5}\mathbf{\Pi}^{3},
\end{eqnarray}
\begin{eqnarray}
\parbox{10mm}{\includegraphics[scale=1.0]{fig10.eps}} = \frac{\mu^{\epsilon}u^{2}}{\epsilon} \left[1 - \frac{1}{2}\epsilon - \frac{1}{2}\epsilon J(P^{2} + K_{\mu\nu}P^{\mu}P^{\nu}) \right]q^{2}\mathbf{\Pi},
\end{eqnarray}
\begin{eqnarray}
\parbox{16mm}{\includegraphics[scale=1.0]{fig11.eps}} = -\frac{\mu^{\epsilon}u^{3}}{\epsilon^{2}} \left[1 - \epsilon - \epsilon J(P^{2} + K_{\mu\nu}P^{\mu}P^{\nu}) \right]q^{4}\mathbf{\Pi}^{2},\quad\quad
\end{eqnarray}
\begin{eqnarray}
\parbox{12mm}{\includegraphics[scale=0.8]{fig21.eps}} = -\frac{\mu^{\epsilon}u^{3}}{2\epsilon^{2}} \left[1 - \frac{1}{2}\epsilon - \epsilon J(P^{2} + K_{\mu\nu}P^{\mu}P^{\nu}) \right]q^{4}\mathbf{\Pi}^2, \quad\quad
\end{eqnarray}
\begin{eqnarray}
&& \parbox{12mm}{\includegraphics[scale=1.0]{fig13.eps}} =   \frac{\mu^{\epsilon}u^{3}}{2\epsilon^{2}} J_{4}(P^{2} + K_{\mu\nu}P^{\mu}P^{\nu})q^{4}\mathbf{\Pi}^{2},
\end{eqnarray}
\begin{eqnarray}
\parbox{10mm}{\includegraphics[scale=1.0]{fig25.eps}} = \frac{3\mu^{\epsilon}u^{3}}{2\epsilon^{2}} \left[1 - \frac{1}{2}\epsilon - \frac{1}{2}\epsilon J(P^{2} + K_{\mu\nu}P^{\mu}P^{\nu}) \right]q^{4}\mathbf{\Pi}^{2},\quad\quad
\end{eqnarray}
\begin{eqnarray}
&& \parbox{12mm}{\includegraphics[scale=1.0]{fig24.eps}} =  -\frac{\mu^{\epsilon}u^{3}}{2\epsilon^{2}} J_{4}(P^{2} + K_{\mu\nu}P^{\mu}P^{\nu})\,q^{4}\mathbf{\Pi}^{2},\quad\quad
\end{eqnarray}
\begin{eqnarray}
&& \parbox{12mm}{\includegraphics[scale=1.0]{fig1.eps}} =
\frac{m^{2}u}{(4\pi)^{2}\epsilon}\left[ 1 - \frac{1}{2}\epsilon\ln\left(\frac{m^{2}}{4\pi\mu^{2}}\right)\right]q\mathbf{\Pi},
\end{eqnarray}
\begin{eqnarray}
\parbox{8mm}{\includegraphics[scale=1.0]{fig2.eps}} = - \frac{m^{2}u^{2}}{(4\pi)^{4}\epsilon^{2}}\left[ 1 - \frac{1}{2}\epsilon - \epsilon\ln\left(\frac{m^{2}}{4\pi\mu^{2}}\right)\right]\,q^{3}\mathbf{\Pi}^{2},
\end{eqnarray}
\begin{eqnarray}
\parbox{12mm}{\includegraphics[scale=1.0]{fig22.eps}} =  \frac{m^{2}g^{2}}{2\epsilon^{2}}\left[ 1 - \frac{1}{2}\epsilon - \frac{1}{2} \epsilon\ln\left(\frac{m^{2}}{4\pi\mu^{2}}\right)\right]q^{3}\mathbf{\Pi}^{2},
\end{eqnarray}
\begin{eqnarray}
&& \parbox{12mm}{\includegraphics[scale=1.0]{fig23.eps}} =  \frac{3m^{2}u^{2}}{2\epsilon^{2}}\left[ 1 - \frac{1}{2} \epsilon\ln\left(\frac{m^{2}}{4\pi\mu^{2}}\right)\right]q^{3}\mathbf{\Pi}^{2},\quad\quad
\end{eqnarray}
\begin{eqnarray}
\left(\parbox{12mm}{\includegraphics[scale=1.0]{fig6.eps}}\right)\Biggr|_{P^{2} + K_{\mu\nu}P^{\mu}P^{\nu}=0} = -\frac{3m^{2}g^{2}}{2\epsilon}\left[ 1 + \frac{1}{2}\epsilon -\epsilon\ln\left(\frac{m^{2}}{4\pi\mu^{2}}\right)\right]q^{3}\mathbf{\Pi}^{2},
\end{eqnarray}
where
\begin{eqnarray}\label{uhduhufgjg}
J(P^{2} + K_{\mu\nu}P^{\mu}P^{\nu}) = \int_{0}^{1}dx \ln \left[\frac{x(1-x)(P^{2}+ K_{\mu\nu}P^{\mu}P^{\nu}) + m^{2}}{\mu^{2}}\right],
\end{eqnarray}
\begin{eqnarray}\label{uhduhufgjgdhg}
&& J_{3}(P^{2} + K_{\mu\nu}P^{\mu}P^{\nu}) = \int_{0}^{1}\int_{0}^{1}dxdy\,(1-y)\times \nonumber \\ && \ln \Biggl\{\frac{y(1-y)(P^{2}+ K_{\mu\nu}P^{\mu}P^{\nu})}{\mu^{2}} + \left[1-y + \frac{y}{x(1-x)}  \right]\frac{m^{2}}{\mu^{2}}\Biggr\},
\end{eqnarray}
\begin{eqnarray}\label{ugujjgdhg}
J_{4}(P^{2} + K_{\mu\nu}P^{\mu}P^{\nu}) = \frac{m^{2}}{\mu^{2}}\int_{0}^{1}dx\frac{(1 - x)}{\frac{x(1 - x)(P^{2} + K_{\mu\nu}P^{\mu}P^{\nu})}{\mu^{2}} + \frac{m^{2}}{\mu^{2}}}.
\end{eqnarray}
Thus we can compute the $\beta_{q}$-function, anomalous dimensions and nontrivial fixed point and obtain 
\begin{eqnarray}\label{rygyfggffgyt}
\beta_{q}(u) = -\epsilon u +   \frac{N + 8}{6}q^{2}\mathbf{\Pi}u^{2} - \frac{3N + 14}{12}q^{4}\mathbf{\Pi}^{2}u^{3} + \frac{N + 2}{36}q^{3}(1-q)\mathbf{\Pi}^{2}u^{3}, 
\end{eqnarray}
\begin{eqnarray}\label{hkfusdrs}
&&\gamma_{\phi,q}(u) = \frac{N + 2}{72}q^{3}\mathbf{\Pi}^{2}u^{2} - \frac{(N + 2)(N + 8)}{1728}q^{5}\mathbf{\Pi}^{3}u^{3},\nonumber \\ &&  
\end{eqnarray}
\begin{eqnarray}\label{kujyhghsghju}
\gamma_{m^{2}, q}(u) = \frac{N + 2}{6}q^{2}\mathbf{\Pi}u -  \frac{5(N + )2}{72}q^{4}\mathbf{\Pi}^{2}u^{2},
\end{eqnarray}
\begin{eqnarray}
u_{q}^{*} = \frac{6\epsilon}{(N + 8)q^{2}\mathbf{\Pi}}\Bigg\{ 1 +  \epsilon\Bigg[ \frac{3(3N + 14)}{(N + 8)^{2}} -\frac{N + 2}{(N + 8)^{2}}\frac{1 - q}{q} \Bigg]\Bigg\},
\end{eqnarray}
where $\mathbf{\Pi} = 1/\sqrt{det(\mathbb{I} + \mathbb{K})}$ is the exact Lorentz-violating full factor \cite{PhysRevD.96.116002}. It was shown early that this factor arises in calculations of quantum corrections in superfield supersymmetric aetherlike Lorentz-breaking models \cite{PhysRevD.86.065035}. If we compute the two independent Lorentz-violating $q$-deformed critical exponents through the two relations $\eta_{q}\equiv \gamma_{\phi,q}(u^{*})$ and $\nu_{q}^{-1}\equiv 2 - \gamma_{m^{2},q}(u^{*})$ and the remaining four ones through the four scaling relations \cite{Stanley}, we obtain the corresponding Lorentz-invariant $q$-deformed ones \cite{PhysRevD.97.105006}, at least at the next-to-leading order considered here. We now proceed to evaluate Lorentz-violating $q$-deformed critical exponents in the next two remaining methods.

\subsection{Callan-Symanzik method}

\par In this and later methods, some trick is used such that the number of Feynman diagrams is reduced to a minimal set of them \cite{Amit}. In the present method, the only needed diagrams are the ones 
\begin{eqnarray} 
\parbox{10mm}{\includegraphics[scale=1.0]{fig10.eps}}_{SP} \equiv \parbox{10mm}{\includegraphics[scale=1.0]{fig10.eps}}\Bigg\vert_{P^{2}=1}, 
\end{eqnarray}
\begin{eqnarray}
\parbox{10mm}{\includegraphics[scale=1.0]{fig6.eps}}^{\prime} \equiv \frac{\partial }{\partial P^{2}}\parbox{12mm}{\includegraphics[scale=1.0]{fig6.eps}}\Bigg\vert_{P^{2}=1},
\end{eqnarray}
\begin{eqnarray}
\parbox{12mm}{\includegraphics[scale=.9]{fig7.eps}}^{\prime} \equiv \frac{\partial }{\partial P^{2}}\parbox{12mm}{\includegraphics[scale=.9]{fig7.eps}}\Bigg\vert_{P^{2}=1},
\end{eqnarray}
\begin{eqnarray} 
\parbox{13mm}{\includegraphics[scale=.9]{fig21.eps}}_{SP} \equiv \parbox{14mm}{\includegraphics[scale=.9]{fig21.eps}}\Bigg\vert_{P^{2}=1}, 
\end{eqnarray}
where $P$ represent the external momenta which are held at fixed values and written in bare mass scale unit whose computed diagrams have the expressions
\begin{eqnarray}
&&\parbox{10mm}{\includegraphics[scale=1.0]{fig10.eps}}_{SP} = \frac{1}{\epsilon}\left(1 - \frac{1}{2}\epsilon \right)q^{2}\mathbf{\Pi},
\end{eqnarray}   
\begin{eqnarray}
&&\parbox{12mm}{\includegraphics[scale=1.0]{fig6.eps}}^{\prime} = -\frac{1}{8\epsilon}\left( 1 - \frac{1}{4}\epsilon +I\epsilon \right)q^{3}\mathbf{\Pi}^{2},
\end{eqnarray}  
\begin{eqnarray}
&&\parbox{12mm}{\includegraphics[scale=0.9]{fig7.eps}}^{\prime} = -\frac{1}{6\epsilon^{2}}\left( 1 - \frac{1}{4}\epsilon +\frac{3}{2}I\epsilon \right)q^{5}\mathbf{\Pi}^{3},
\end{eqnarray}  
\begin{eqnarray}
&&\parbox{12mm}{\includegraphics[scale=0.8]{fig21.eps}}_{SP} = \frac{1}{2\epsilon^{2}}\left(1 - \frac{1}{2}\epsilon \right)q^{4}\mathbf{\Pi}^{2},
\end{eqnarray}  
where the integral $I$ \cite{PhysRevD.8.434}
\begin{eqnarray}
&& I = \int_{0}^{1} \left\{ \frac{1}{1 - x(1 - x)} + \frac{x(1 - x)}{[1 - x(1 - x)]^{2}}\right\}
\end{eqnarray}
is a number and is due to the symmetry point chosen. Thus we obtain 
\begin{eqnarray}\label{fhdfghg}
\beta_{q}(u) = -\epsilon u +   \frac{N + 8}{6}\left( 1 - \frac{1}{2}\epsilon \right)q^{2}\mathbf{\Pi}u^{2} - \frac{3N + 14}{12}q^{4}\mathbf{\Pi}^{2}u^{3} + \frac{N + 2}{36}q^{3}\mathbf{\Pi}^{2}(1-q)u^{3}, 
\end{eqnarray}
\begin{eqnarray}
\gamma_{\phi ,q} = \frac{N + 2}{72}\left( 1 - \frac{1}{4}\epsilon + I\epsilon \right)q^{3}\mathbf{\Pi}^{2}u^{2} -  \frac{(N + 2)(N + 8)}{432}\left( 1 + I \right)q^{5}\mathbf{\Pi}^{3}u^{3},  
\end{eqnarray}
\begin{eqnarray}\label{sfdtsvdb}
&&\overline{\gamma}_{\phi^{2}, q}(u) = \frac{N + 2}{6}\left( 1 - \frac{1}{2}\epsilon \right)q^{2}\mathbf{\Pi}u -  \frac{N + 2}{12}q^{4}\mathbf{\Pi}^{2}u^{2},\nonumber \\ &&
\end{eqnarray}
\begin{eqnarray}
u_{q}^{*} = \frac{6\epsilon}{(N + 8)q^{2}\mathbf{\Pi}}\Bigg\{ 1 +  \epsilon\Bigg[ \frac{3(3N + 14)}{(N + 8)^{2}} + \frac{1}{2} -\frac{N + 2}{(N + 8)^{2}}\frac{1 - q}{q} \Bigg]\Bigg\},
\end{eqnarray}
where $\overline{\gamma}_{\phi^{2},q}(u) = \gamma_{\phi^{2},q}(u) - \gamma_{\phi,q}(u)$. The integral $I$ is a number that can be evaluated analytically in terms of the dilogarithm function of certain argument \cite{Ramond}. Now by applying the relations $\eta_{q}\equiv\gamma_{\phi,q}(u^{*})$ and $\nu_{q}^{-1}\equiv 2 - \eta_{q} - \overline{\gamma}_{\phi^{2},q}(u^{*})$, the integral $I$ is is canceled out in the middle of computations and we obtain that the Lorentz-violating $q$-deformed critical exponents are the same as their Lorentz-invariant counterparts \cite{PhysRevD.97.105006}.

\subsection{Unconventional minimal subtraction scheme}
     
\par In this method \cite{J.Math.Phys.542013093301}, the minimal set of Feynman diagrams is the one 
\begin{eqnarray} 
\parbox{11mm}{\includegraphics[scale=1.0]{fig10.eps}}, 
\end{eqnarray}
\begin{eqnarray}
\parbox{11mm}{\includegraphics[scale=1.0]{fig6.eps}},
\end{eqnarray}
\begin{eqnarray}
\parbox{11mm}{\includegraphics[scale=.9]{fig7.eps}},
\end{eqnarray}
\begin{eqnarray} 
\parbox{13mm}{\includegraphics[scale=.9]{fig21.eps}}, 
\end{eqnarray}
respectively, where now the external momenta can assume arbitrary values. With the expressions for the calculated diagrams given by
\begin{eqnarray}
\parbox{10mm}{\includegraphics[scale=1.0]{fig10.eps}} = \frac{1}{\epsilon} \left[1 - \frac{1}{2}\epsilon - \frac{1}{2}\epsilon L(P^{2} + K_{\mu\nu}P^{\mu}P^{\nu}, m_{B}^{2}) \right]q^{2}\mathbf{\Pi},
\end{eqnarray}   

\begin{eqnarray}
&&\parbox{12mm}{\includegraphics[scale=1.0]{fig6.eps}} = \left\{-\frac{3 m_{B}^{2}}{2 \epsilon^{2}}\left[1 + \frac{1}{2}\epsilon + \left(\frac{\pi^{2}}{12} +1 \right)\epsilon^{2} \right] -  \frac{3 m_{B}^{2}}{4}\tilde{i}(P^{2} + K_{\mu\nu}P^{\mu}P^{\nu}, m_{B}^{2})  - \right.  \nonumber \\  &&\left. \frac{(P^{2} + K_{\mu\nu}P^{\mu}P^{\nu})}{8 \epsilon}\left[1 + \frac{1}{4}\epsilon - 2 \epsilon L_{3}(P^{2} + K_{\mu\nu}P^{\mu}P^{\nu},m_{B}^{2})\right]\right\}q^{3}\mathbf{\Pi}^{2}, 
\end{eqnarray}
\begin{eqnarray}
&&\parbox{12mm}{\includegraphics[scale=1.0]{fig7.eps}} = \left\{-\frac{5 m_{B}^{2}}{3 \epsilon^{3}}\left[1 + \epsilon + \left(\frac{\pi^{2}}{24} + \frac{15}{4} \right)\epsilon^{2} \right] -  \frac{5 m_{B}^{2}}{2 \epsilon}\tilde{i}(P^{2} + K_{\mu\nu}P^{\mu}P^{\nu}, m_{B}^{2}) \right.  \nonumber \\  &&\left. -\frac{P^{2} + K_{\mu\nu}P^{\mu}P^{\nu}}{6 \epsilon^{2}}\left[1+ \frac{1}{2}\epsilon - 3 \epsilon L_{3}(P^{2} + K_{\mu\nu}P^{\mu}P^{\nu},m_{B}^{2})\right]\right\}q^{5}\mathbf{\Pi}^{3}, 
\end{eqnarray}
\begin{eqnarray}
\parbox{14mm}{\includegraphics[scale=1.0]{fig21.eps}} = \frac{1}{\epsilon^{2}} \left[1 - \frac{1}{2}\epsilon - \epsilon L(P^{2} + K_{\mu\nu}P^{\mu}P^{\nu}, m_{B}^{2}) \right]q^{4}\mathbf{\Pi}^{2},
\end{eqnarray}  
where
\begin{eqnarray}
L(P^{2} + K_{\mu\nu}P^{\mu}P^{\nu}, m_{B}^{2}) = \int_{0}^{1}dx\ln[x(1-x)(P^{2} + K_{\mu\nu}P^{\mu}P^{\nu}) + m_{B}^{2}],
\end{eqnarray}
\begin{eqnarray}
L_{3}(P^{2} + K_{\mu\nu}P^{\mu}P^{\nu}, m_{B}^{2}) = \int_{0}^{1}dx(1-x) \ln[x(1-x)(P^{2} + K_{\mu\nu}P^{\mu}P^{\nu}) + m_{B}^{2}],
\end{eqnarray}
\begin{eqnarray}
&&\tilde{i}(P^{2} + K_{\mu\nu}P^{\mu}P^{\nu}, m_{B}^{2}) = \int_{0}^{1} dx \int_{0}^{1}dy\ln y \times \nonumber \\ &&   \frac{d}{dy}\left((1-y)\ln\left\{y(1-y)P^{2} + \left[1-y + \frac{y}{x(1-x)}\right]m_{B}^{2} \right\}\right),
\end{eqnarray}
we obtain for the Lorentz-violating $q$-deformed $\beta_{q}$-function, anomalous dimensions and nontrivial fixed point
\begin{eqnarray}\label{rygyfggfgfgyt}
\beta_{q}(u) = -\epsilon u +   \frac{N + 8}{6}q^{2}\mathbf{\Pi}u^{2} -  \frac{3N + 14}{12}q^{4}\mathbf{\Pi}^{2}u^{3} + \frac{N + 2}{36}q^{3}(1-q)\mathbf{\Pi}^{2}u^{3}, 
\end{eqnarray}
\begin{eqnarray}\label{hkfdusdrs}
&&\gamma_{\phi,q}(u) = \frac{N + 2}{72}q^{3}\mathbf{\Pi}^{2}u^{2} - \frac{(N + 2)(N + 8)}{1728}q^{5}\mathbf{\Pi}^{3}u^{3},\nonumber \\ &&  
\end{eqnarray}
\begin{eqnarray}\label{sfdtsvdb}
\overline{\gamma}_{\phi^{2}, q}(u) = \frac{N + 2}{6}q^{2}\mathbf{\Pi}u -  \frac{N + 2}{12}q^{4}\mathbf{\Pi}^{2}u^{2},
\end{eqnarray}
\begin{eqnarray}
u_{q}^{*} = \frac{6\epsilon}{(N + 8)q^{2}\mathbf{\Pi}}\Bigg\{ 1 +  \epsilon\Bigg[ \frac{3(3N + 14)}{(N + 8)^{2}} -\frac{N + 2}{(N + 8)^{2}}\frac{1 - q}{q} \Bigg]\Bigg\}.
\end{eqnarray}
Once again, the relations $\eta_{q}\equiv\gamma_{\phi,q}(u^{*})$ and $\nu_{q}^{-1}\equiv 2 - \eta_{q} - \overline{\gamma}_{\phi^{2},q}(u^{*})$ when applied by using the Lorentz-violating $q$-deformed nontrivial fixed point aforementioned we obtain the Lorentz-invariant $q$-deformed critical exponents \cite{PhysRevD.97.105006}. This completes our proposed task and shows that the $q$-deformed critical exponents besides being universal quantities since are the same in the three distinct and independent methods are insensible to the Lorentz-violating symmetry breaking mechanism. This confirms the universality hypothesis, at least at the next-to-leading order approached here, once the referred symmetry breaking mechanism occur in the spacetime where the field is defined and not in its internal one. Now we generalize our results for any loop level.

\section{All-loop order Lorentz-violating $q$-deformed critical exponents}

\par In this Sect. we have to generalize the next-to-leading order Lorentz-violating $q$-deformed critical exponents results for any loop level. For that, we have to apply a general theorem \cite{PhysRevD.96.116002}. We have to employ the BPHZ method, since the critical exponents are universal quantities. The referred theorem asserts that a general Feynman diagram of any loop level, when computed, assumes the expression $\mathbf{\Pi}^{L}\mathcal{F}_{q}(u,P^{2} + K_{\mu\nu}P^{\mu}P^{\nu},\epsilon,\mu)$ if its evaluated Lorentz-invariant counterpart is given by $\mathcal{F}(u,P^{2},\epsilon,\mu)$, where $L$ is the its loop number. According to the BPHZ prescription, the momentum-dependent integrals cancel out in the middle of calculation and we can write for all-loop order Lorentz-violating $q$-deformed $\beta_{q}$-function and anomalous dimensions for all-loop order as
\begin{eqnarray}\label{uhgufhduhufdhu}
\beta_{q}(u) =  -\epsilon u + \sum_{n=2}^{\infty}\beta_{q,n}^{(0)}\mathbf{\Pi}^{n-1}u^{n}, 
\end{eqnarray}
\begin{eqnarray}
\gamma_{\phi,q}(u) = \sum_{n=2}^{\infty}\gamma_{q,n}^{(0)}\mathbf{\Pi}^{n}u^{n},
\end{eqnarray}
\begin{eqnarray}
\gamma_{m^{2},q}(u) = \sum_{n=1}^{\infty}\gamma_{m^{2},q,n}^{(0)}\mathbf{\Pi}^{n}u^{n},
\end{eqnarray}
where $\beta_{q,n}^{(0)}$, $\gamma_{\phi,q,n}^{(0)}$ and $\gamma_{\phi^{2},q,n}^{(0)}$ are Lorentz-invariant $q$-deformed factors. From the all-loop order Lorentz-violating $q$-deformed $\beta_{q}$-function aforementioned, we can evaluate the all-loop Lorentz-violating $q$-deformed nontrivial fixed point. Its value is given by $u_{q}^{*} = u_{q}^{*(0)}/\mathbf{\Pi}$, where $u_{q}^{*(0)}$ is the Lorentz-invariant $q$-deformed nontrivial fixed point for all loop orders since we can factor one power of $u$ from $\beta_{q}$. Then we obtain that the Lorentz-violating $q$-deformed critical exponents at all-loop levels present the same values as their Lorentz-invariant $q$-deformed counterparts. Thus we have attained our goal initially proposed: that of confirming the universality hypothesis by showing that the Lorentz-violating mechanism, even treated exactly, \emph{i. e.} for all loop orders in the Lorentz-violating parameters $K_{\mu\nu}$, does not effect the Lorentz-violating $q$-deformed critical exponents values once the referred exact symmetry breaking mechanism is one occurring not in the internal space of the $q$-deformed field but in the spacetime where it is embedded. Now we proceed to present our conclusions.

\section{Conclusions}\label{Conclusions}

\par In this work we have probed analytically the influence of exact Lorentz-violating mechanism on the next-to-leading order radiative quantum corrections to critical exponents of massive $q$-deformed O($N$) $\lambda\phi^{4}$ scalar field theories. After that, through an induction process, we have generalized our results for any loop level by applying a general theorem. We have performed the calculations through three distinct and independent field-theoretic renormalization group methods and obtained that the values of the corresponding critical exponents were the same when computed through that three methods. The tensor $K_{\mu\nu}$ is not renormalized, because to remove the divergences of the theory it is sufficient to renormalize the field, mass and coupling constant. We have found that the exact Lorentz-violating mechanism has not affected the all-loop $q$-deformed critical exponents values. This fact confirms the universality hypothesis since the referred exact symmetry breaking mechanism occurs in the spacetime where the field is defined and not in its internal one.

\section*{Acknowledgments}
PRSC and MISJ would like to thank CAPES (brazilian funding agency) for financial support. PRSC would like to
thank CNPq (brazilian funding agency) through grant Universal-431727/2018.

\bibliography{apstemplate}

\end{document}